\documentclass[jasms]{ametsoc}          % For manuscript

\usepackage{amsfonts}                                                    % Standard CM fonts.
\usepackage{amsmath}                                                     % Standard CM fonts.
\usepackage{amsbsy}                                                      % Standard CM fonts.
\usepackage{graphicx}
\newcommand{\D}{\mathrm{D}}

\newcommand{\jet}{\mathrm{jet}}
\renewcommand{\d}{\mathrm{d}}

\lefthead{BERON-VERA, BROWN, OLASCOAGA, RYPINA, KO\c{C}AK AND UDOVYDCHENKOV}%
\righthead{ACCEPTED FOR PUBLICATION IN THE JOURNAL OF THE ATMOSPHERIC SCIENCES}%
\received{\today}

\sectionnumbers
\printfigures

\authoraddr{F.~J. Beron-Vera,
RSMAS/AMP, University of Miami,
4600 Rickenbacker Cswy.,
Miami, FL 33149
(fberon@rsmas.miami.edu)\\
$^{*}$Current affiliation: Woods Hole Oceanographic Institution, Woods Hole, Massachusetts.}

\slugcomment{Accepted for publication in the \emph{Journal of the Atmospheric
Sciences}.\\
(Started: April 13, 2007; this version: March 17, 2008.)}

\begin{document}

\title{Zonal Jets as Transport Barriers in Planetary
Atmospheres}

\author{F.~J. Beron-Vera, M.~G. Brown, M.~J. Olascoaga and I.~I. Rypina$^{*}$}
\affil{Rosenstiel School of Marine and Atmospheric Science,
University of Miami, Miami, Florida}
\author{H. Ko\c{c}ak}
\affil{Departments of Computer Science and Mathematics,
University of Miami, Coral Gables, Florida}
\author{I.~A. Udovydchenkov$^{*}$}
\affil{Rosenstiel School of Marine and Atmospheric Science,
University of Miami, Miami, Florida}

\begin{abstract}
The connection between transport barriers and potential
vorticity (PV) barriers in PV-conserving flows is investigated
with a focus on zonal jets in planetary atmospheres. A
perturbed PV-staircase model is used to illustrate important
concepts. This flow consists of a sequence of narrow eastward
and broad westward zonal jets with a staircase PV structure;
the PV-steps are at the latitudes of the cores of the eastward
jets. Numerically simulated solutions to the quasigeostrophic
PV conservation equation in a perturbed PV-staircase flow are
presented. These simulations reveal that both eastward and
westward zonal jets serve as robust meridional transport
barriers. The surprise is that westward jets, across which the
background PV gradient vanishes, serve as robust transport
barriers. A theoretical explanation of the underlying barrier
mechanism is provided. It is argued that transport barriers
near the cores of westward zonal jets, across which the
background PV gradient is small, are found in Jupiter's
midlatitude weather layer and in the Earth's summer hemisphere
subtropical stratosphere.
\end{abstract}

\begin{article}

\section{Introduction}

In \citet{Rypina-etal-07a} it was argued that the transport
barrier near the core of the austral polar night jet can be
explained by a mechanism different from the potential vorticity
(PV) barrier mechanism \citep{Juckes-McIntyre-87}. The new
barrier mechanism, which was subsequently referred to as
``strong KAM stability'' \citep{Rypina-etal-07b}, follows from
an argument that does not make use of dynamical constraints on
the streamfunction. This necessitates that dynamical
constraints be considered separately. Interestingly, decoupling
of the dynamical constraints from the barrier mechanism leads
to the possibility that transport barriers in PV-conserving
flows may occur at locations that do not coincide with
PV-barriers. \citet{Rypina-etal-07a} predicted that barriers of
this type should be present in close proximity to the cores of
westward zonal jets in planetary atmospheres. In this paper we
demonstrate that transport barriers of this type are present in
a numerically simulated PV-conserving flow. We also argue that
barriers of the type described are present in Jupiter's
midlatitude weather layer and in the Earth's summer hemisphere
subtropical stratosphere.

In the following section passive tracer transport in a
numerically simulated perturbed PV-staircase flow is
investigated. It is shown that robust meridional transport
barriers in close proximity to the cores of both eastward and
westward zonal jets are present. The surprise is that westward
jets, at which the background PV-gradient vanishes, act as
transport barriers. Essential elements of the strong KAM
stability argument are reviewed to explain this behavior. In
section 3 we discuss the relevance of transport barriers of the
strong KAM stability type to: 1) the interpretation of
Jupiter's midlatitude weather layer belt-zone structure; and 2)
the Earth's summer hemisphere subtropical stratosphere. In the
final section we briefly discuss our results.

\section{Transport barriers in a perturbed PV-staircase flow}

\begin{figure}[tb]
\figbox*{}{}{
\includegraphics[width=.5\textwidth,clip=]{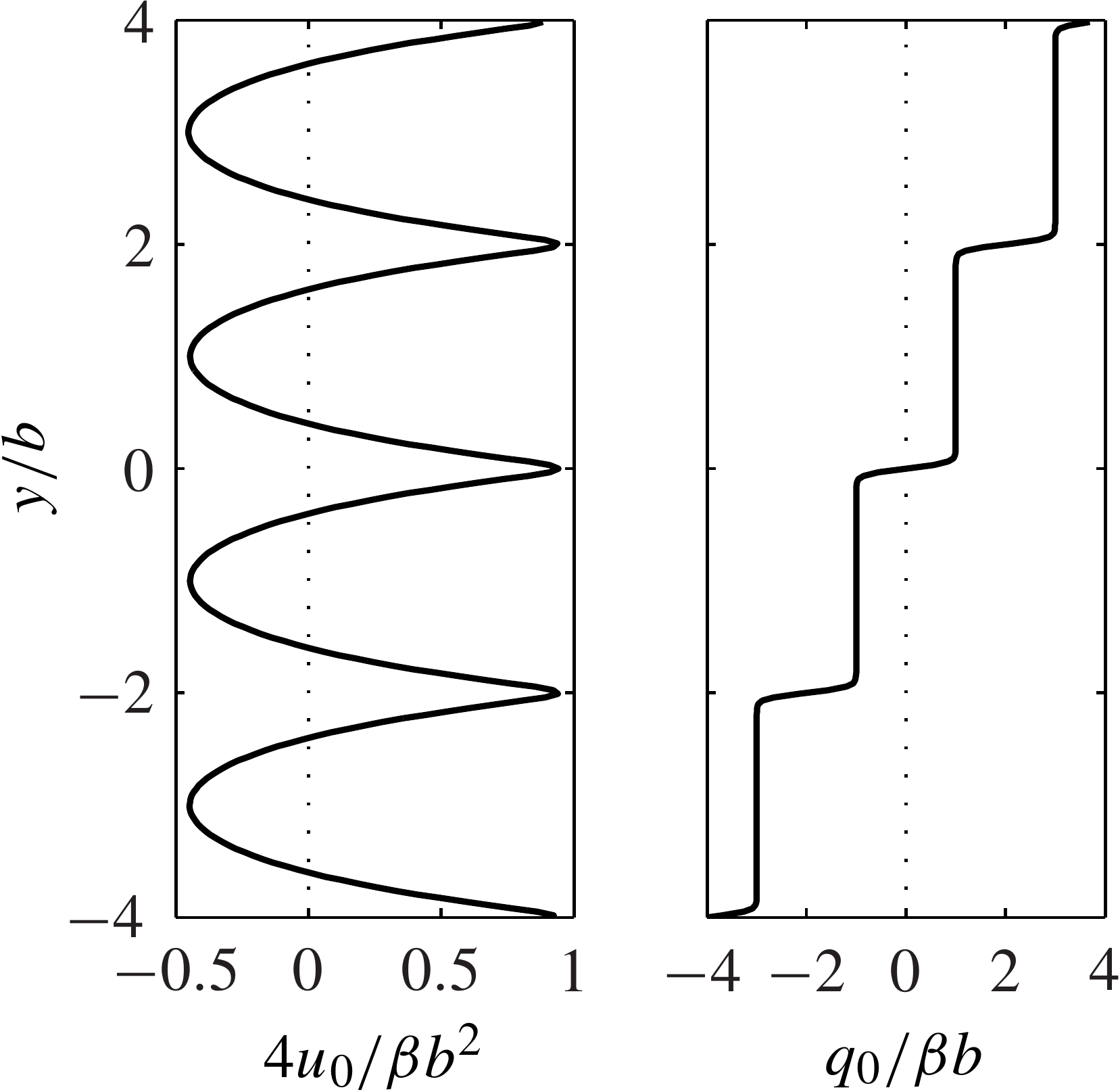}}
\caption{Zonal velocity (left panel) and potential vorticity
(right panel) in a PV-staircase zonal flow. The zonal velocity
structure shown corresponds to the finite $L_\D$ case, Eq.
(\ref{u_0-1}), with $L_\D = b/2 = $ 2000 km, roughly
approximating midlatitude conditions on Jupiter. In this figure
the ideal PV jumps have smooth transition regions with tanh
dependence on $y$ locally.} \label{u0q0}
\end{figure}

In this section we consider passive tracer transport in a
perturbed PV-staircase flow.  We assume quasigeostrophic dynamics
in a one-layer reduced-gravity setting and make use of a local
Cartesian coordinate system $(x,y)$ where $x$ and $y$ increase to
the east and north, respectively, and the constant $\beta$ is the
local $y$-derivative of the Coriolis parameter. The zonal and
meridional components of the velocity field are $u = -
\partial \psi/\partial y$ and $v =
\partial \psi/\partial x$, respectively, where $\psi(x,y,t)$ is
the streamfunction. The flow is constrained to satisfy
\begin{eqnarray}\label{qcons}
    \frac{\partial q}{\partial t} - \frac{\partial
    \psi}{\partial y} \frac{\partial q}{\partial x} +
    \frac{\partial \psi}{\partial x} \frac{\partial q}
    {\partial y} = 0
\end{eqnarray}
where
\begin{eqnarray}\label{qdef}
    q = \nabla^2 \psi - L_\D^{-2} \psi + \beta y
\end{eqnarray}
is the quasigeostrophic potential vorticity and $L_\D$ is the
deformation radius. Recent theoretical, numerical and experimental
work, including extensions involving spherical geometry,
shallow-water dynamics and inclusion of weak forcing and
dissipation, has shown that flows satisfying
(\ref{qcons})--(\ref{qdef}) with periodic boundary conditions in
$x$ tend to evolve toward a state of the form
\begin{eqnarray}\label{psi}
    \psi = \psi_0(y) + \psi_1(x,y,t)
\end{eqnarray}
where $\psi_1$ is a small perturbation to $\psi_0$
\citep{Rhines-75,Vallis-Maltrud-93,
Manfroi-Young-99,Danilov-Gryanik-04,Danilov-Gurarie-04,Dritschel-McIntyre-08,
Williams-78,Nozawa-Yoden-97,Huang-Robinson-98,Cho-Polvani-96,
Peltier-Stunhe-02,Scott-Polvani-07,Read-etal-07}. The
background zonal flow is characterized by an approximately
piecewise constant PV distribution that has been appropriately
described as a PV-staircase
\citep{Dritschel-McIntyre-08,Baldwin-etal-07,Dunkerton-Scott-07}.
The corresponding zonal velocity profile $u_0(y) = - \d
\psi_0/\d y$ is periodic in $y$. Taking the period to be $2b$,
the jump in $q_0(y)$ at each step is $2b\beta$, and the zonal
flow is the periodic extension of
\begin{eqnarray}\label{u_0-1}
    u_0(y) = \beta L_\D^2 \left(\frac{b}{L_\D}
    \frac{\cosh ((y-b)/L_\D)}{\sinh (b/L_\D)} - 1 \right),
    \quad 0 \leq y \leq 2b,
\end{eqnarray}
consisting of a periodic sequence of alternating narrow
eastward and broad westward zonal jets with $q_0(y)$ piecewise
constant between adjacent eastward jets, as illustrated in Fig.
\ref{u0q0}. Note that at the center of each constant $q_0$ band
lies a westward jet. In the limit $L_\D/b \rightarrow \infty$
(\ref{u_0-1}) reduces to
\begin{eqnarray}\label{u_0-2}
    u_0(y) = \frac{\beta}{2}\left((y-b)^2 -
    \frac{b^2}{3}\right),
    \quad 0 \leq y \leq 2b,
\end{eqnarray}
whose qualitative features are identical to those of the finite
$L_\D/b$ case.  The flows (\ref{u_0-1}) and (\ref{u_0-2}) are
normalized so that the integral of $u_0(y)$ over $y$ vanishes. The
Rhines scale $\sqrt{U/\beta}$, where $U$ is a characteristic
velocity, is an approximate measure of the separation between
adjacent eastward jets. Corresponding to (\ref{u_0-2}) this
separation is exactly $2b = 2\sqrt{3U_{\max}/\beta}$ where
$U_{\max}$ is the wind speed at the core of one of the eastward
jets.

Before presenting numerical simulations of passive tracer
transport in a perturbed PV-staircase flow we describe
predictions based on two different arguments of the expected
locations of transport barriers in this flow.  First, the
PV-barrier argument \citep{Dritschel-McIntyre-08} leads to the
expectation that transport barriers should be present only near
the cores of the eastward jets.  The basic elements of the
PV-barrier argument were used originally by
\citet{Juckes-McIntyre-87} to explain the mechanism by which
the eastward jet at the perimeter of the austral stratospheric
polar vortex, sometimes referred to as the austral polar night
jet, during the late winter and early spring serves to trap
ozone-depleted air inside the polar vortex. The essential
elements of the argument are that at eastward jets the large
gradient of $q_0(y)$ is associated with a large Rossby wave
restoring force (``Rossby wave elasticity'') which inhibits
meridional exchange of fluid at larger scales and that shear
$u_0'(y) = \d u_0 /\d y$ acts to inhibit meridional exchange at
smaller scales. (But note that in \citet{Rypina-etal-07a} it is
argued that increasing meridional shear acts, on average, to
increase meridional exchange.)

An alternative argument, based on the strong KAM stability
mechanism \citep{Rypina-etal-07a,Rypina-etal-07b}, leads to the
expectation that transport barriers should be present near the
cores of both eastward and westward jets in a PV-staircase flow.
The argument leading to this expectation will now be reviewed.
The Lagrangian (particle trajectory) equations of motion,
\begin{eqnarray}\label{Lagrange}
    \frac{\d x}{\d t} = -\frac{\partial \psi}{\partial y},\quad
    \frac{\d y}{\d t} = \frac{\partial \psi}{\partial x},
\end{eqnarray}
constitute a nonautonomous one-degree-of-freedom Hamiltonian
system, with $(x,y)$ the canonically conjugate
coordinate--momentum pair and $\psi(x,y,t)$ the Hamiltonian.
This allows results from studies of integrable and
nonintegrable Hamiltonian systems to be applied. In the
background steady flow, with $\psi = \psi_0(y)$, equations
(\ref{Lagrange}) are integrable and the motion is describable
using a transformed Hamiltonian $H_0(I)$ where $(I,\theta)$ are
action--angle variables. Each trajectory lies on a torus which
is labeled by its $I$-value. Motion is $2\pi$-periodic in
$\theta$ with angular frequency $\omega(I) = H'_0(I)$.
According to each of many variants of the
Kolmogorov--Arnold--Moser (KAM) theorem \citep{Arnold-etal-86},
many of the unperturbed tori survive in the perturbed system
(\ref{psi}), albeit in a slightly distorted form, provided
certain conditions are met. Surviving tori cannot be traversed
and serve as transport barriers. (For reasons described in
\citep{Rypina-etal-07a}, the process known as Arnold diffusion
does not occur in the systems under study.) Torus destruction
is caused by the excitation and overlapping of resonances. Each
resonance has a characteristic width $\Delta \omega$.
Nondegenerate $\omega'(I) \neq 0$ resonance widths are
proportional to $|\omega'(I)|^{1/2}$. Degenerate $\omega'(I) =
0$ resonance widths do not vanish but are generally narrower
than nondegenerate resonance widths. (Quantitative estimates of
both degenerate and nondegenerate resonance widths are given in
\citep{Rypina-etal-07b}; for our purposes it suffices to note
the general trend.)  For most moderate strength perturbations
to the background, small resonance widths near degenerate tori
lead to nonoverlapping resonances and thus unbroken tori that
serve as transport barriers. This constitutes the strong KAM
stability barrier argument. In our model (\ref{psi}) the
connection between $(x,y)$ and $(I,\theta)$ is particularly
simple: $I = -Ry/(2\pi)$, $\theta = 2\pi x/R$ where $R$ is the
distance around the planet along a constant latitude circle at
the latitude at which $\beta$ is defined. The period of motion
$2\pi/\omega$ is $R/u_0(y)$, so $\omega(I) = 2\pi
R^{-1}u_0(-I/R)$. At the cores of both eastward and westward
jets $u'_0(y) = 0$, so $\omega'(I) = 0$ at these locations and
the strong KAM stability argument predicts that robust
meridional transport barriers should be present. Barriers of
this type may be broken if the transient perturbation
$\psi_1(x,y,t)$ strongly excites a low-order resonance with the
frequency $\omega = 2\pi u_0(y)/R$ of the background flow near
the core of the jet. Three final issues are noteworthy. First,
because the strong KAM stability argument is a kinematic
argument (based on (\ref{Lagrange}) alone), dynamical
consistency---consistency between (\ref{qcons})--(\ref{psi})
and (\ref{u_0-1}) or (\ref{u_0-2})---and flow stability must be
considered separately. Second, our emphasis on jets is
unnecessarily restrictive inasmuch as the strong KAM stability
argument holds at all locations where $u_0'(y) = 0$. Third, the
stated results on KAM theory assume that $\psi_1(x,y,t)$ can be
expressed as a multiperiodic (generically quasiperiodic)
function of $t$.

We now describe a set of numerical experiments that were
performed to investigate passive tracer transport in a
perturbed PV-staircase flow.  The streamfunctions on which our
tracer transport simulations are based were constructed by
numerically solving (\ref{qcons})--(\ref{qdef}) using as an
initial state a perturbation to the background PV-staircase
(\ref{u_0-1}). Before presenting the results, it is appropriate
to make two comments about what we expect to learn from these
simulations. First, these simulations provide a test of the
assertion that the decomposition (\ref{qcons})--(\ref{u_0-1})
is dynamically consistent. Second, given a positive outcome of
the first test, these simulations test whether westward zonal
jets, across which there is no PV-barrier, serve as robust
meridional transport barriers.  The quasigeostrophic equation
(\ref{qcons})--(\ref{qdef}) was solved numerically using a
standard pseudospectral technique on a $256^2$ grid in the
$[0,8b) \times [-4b,4b)$ computational domain. Periodic
boundary conditions in $\psi$ were applied in both the $x$ and
$y$ directions. The Arakawa representation of the advective
terms, which are often written as a Jacobian, was used. The
solution was marched forward in time using a second-order
Adams--Bashforth scheme with a dimensionless timestep $\beta
b\Delta t = 0.002$. To control the spurious amplification of
high wavenumber modes, we applied a weak exponential cutoff
filter and included a small amount of hyperviscosity. Two types
of initial perturbation to the background PV-staircase were
used in our simulations. The first type of perturbation
consisted of a superposition of periodic displacements of PV
contours with random phases uniformly distributed on
$[0,2\pi)$. The second type of perturbation was a doubly
periodic perturbation to the streamfunction consisting of a sum
of a product of Fourier modes with random phases. In all of our
simulations the separation between adjacent eastward jets $2b$
was taken to be 8000 km and $\beta = 3.442 \times 10^{-9}$
km$^{-1}$ s$^{-1}$ was used. The simulations shown correspond
to $L_\D = b/2$. In anticipation of our discussion of Jupiter
in the following section, these parameters were chosen to
approximately reproduce conditions on Jupiter. Note, however,
that in our simulations the period in $x$ is 32000 km, which is
approximately one-tenth the midlatitude distance around Jupiter
on a line of constant latitude. Many one-year model simulations
were run. For some parameter values ten-year model simulations
were performed. For the parameter values given, the change in
both energy and enstrophy throughout the duration of the
simulations performed was less than $1\%$, giving us confidence
in the accuracy of the simulations. Both types of initial
perturbation gave similar results.

\begin{figure}[tb]
\figbox*{}{}{
\includegraphics[width=\textwidth,clip=]{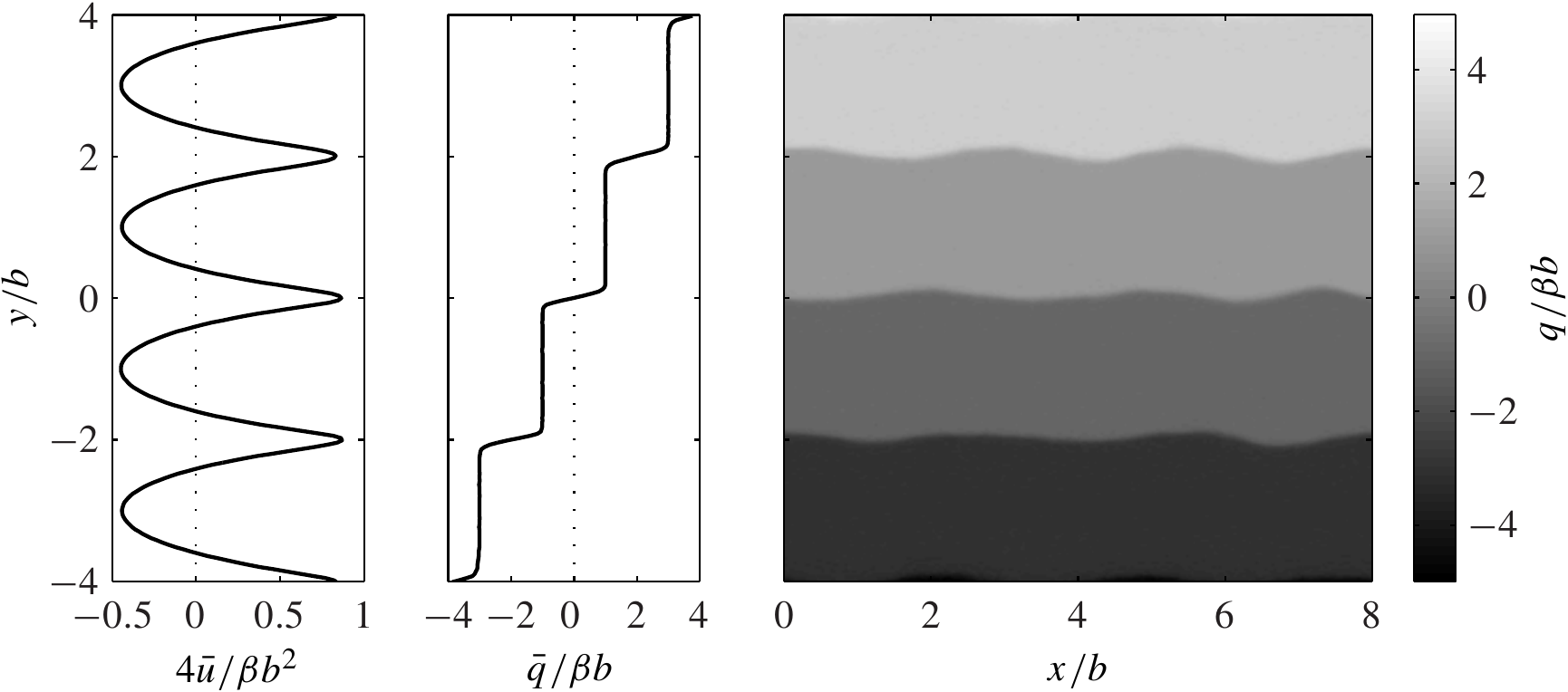}}
\caption{Instantaneous zonally-averaged zonal velocity (left
panel), zonally-averaged potential vorticity (middle panel),
and potential vorticity (right panel) after a one-year
simulation of the quasigeostrophic equation
(\ref{qcons})--(\ref{qdef}) using as initial state a
perturbation to the PV-staircase flow of Fig. \ref{u0q0}.}
\label{uq}
\end{figure}

\begin{figure}[tb]
\figbox*{}{}{
\includegraphics[width=\textwidth,clip=]{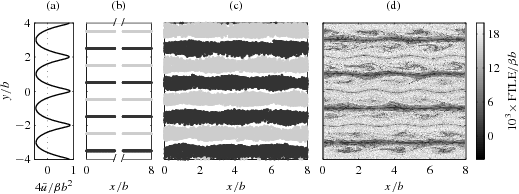}}
\caption{(a) Zonally averaged zonal velocity $\bar{u}(y)$ at $t =
1$ year. (b) Initial positions $(x(0), y(0))$ of color-coded
passive tracers.  (c) Positions $(x(\tau),y(\tau))$ of color coded
passive tracers at $t = \tau = 1$ year . (d) Finite-time Lyapunov
exponent (FTLE) field as a function of initial position computed
using a one year integration time interval. The time-dependent
velocity field whose final state is shown in Fig. \ref{uq} was
used to construct (a), (c) and (d).} \label{FTLE}
\end{figure}

Figure \ref{uq} shows plots of instantaneous, at $ t = \tau =
1$ year, zonally-averaged zonal velocity $\bar{u}(y,\tau)$,
zonally averaged potential vorticity $\bar{q}(y,\tau)$ and
potential vorticity $q(x,y,\tau)$. Comparison of Figs.
\ref{u0q0} and \ref{uq} provides strong support for the
dynamical consistency of the decomposition
(\ref{psi})--(\ref{u_0-1}), but these plots provide no insight
into whether transport barriers are present. To address the
latter question we have used the year-long records of computed
velocity fields to: 1) follow the evolution of distributions of
passive tracers, which evolve according to (\ref{Lagrange});
and 2) compute finite-time Lyapunov exponents \citep[FTLEs;
see, for example,][]{Haller-01b}. Typical results are shown in
Fig. \ref{FTLE}. As shown in the figure, the initial positions
of the passive tracers fell on the lines $\pm 2y/b = \{1, 3, 5,
7\}$, which lie midway between the unperturbed eastward and
westward jets. It is seen that after a one-year integration the
regions between jets are well-mixed, but that there is no
meridional transport across the cores of the zonal jets. It
should be emphasized that in this and other simulations tracer
particles spread meridionally to fill the domains shown in
about two weeks. Throughout the reminder of the one-year
integration no additional meridional tracer spreading occurs.
With this in mind, Fig. \ref{FTLE} clearly shows that both
eastward and westward zonal jets act as meridional transport
barriers, consistent with the strong KAM stability argument.
Calculation of FTLEs provides an additional test of the
correctness of the strong KAM stability argument. Lyapunov
exponents are a measure of the rate of divergence of
neighboring trajectories. According to the strong KAM stability
argument, the transport barriers at the cores of zonal jets
coincide with generally thin bands of KAM invariant tori on
which Lyapunov exponents are zero. Finite time estimates of
Lyapunov exponents will not be identically zero on KAM
invariant tori, but these structures should be readily
identifiable as thin bands of low FTLE estimates. This is
precisely what is seen in Fig. \ref{FTLE}; both westward and
eastward jets are identifiable as thin bands of low FTLE
estimates, consistent with the strong KAM stability argument.
Typical computed values of FTLE, in units of $10^{-3}\beta b$,
shown in Fig. \ref{FTLE} are 2 at the cores of the westward
jets, 7 at the cores of the eastward jets, and 17 in the
well-mixed regions.

We have performed many numerical experiments based on
PV-staircase flows of the type described here. These
simulations support the conclusion that both eastward zonal
jets (where $\partial \bar{q}/\partial y$ is very large) and
westward zonal jets (where $\partial \bar{q}/\partial y$ is
very small) act as robust meridional transport barriers. This
conclusion is not sensitive to the choice of parameter values
or details of the initial perturbation. Two general trends are
noteworthy. First, for a small perturbation the width of the
barrier region near westward jets is greater than the width of
the barrier region near eastward jets. This behavior is
consistent with the strong KAM stability argument:
$|\omega'(I)|$ is small over a larger $y$-domain near westward
jets than near eastward jets. Second, as the perturbation
strength increases, transport barriers near westward jets
generally break before eastward jet barriers break. In our
simulations the westward jet barriers broke when the initial
rms meridional PV-contour displacement exceeded approximately
$b/4$, while the eastward jet barriers broke when the initial
rms displacement was approximately twice this value. Possible
explanations for the somewhat more robust nature of the
eastward jets are: 1) the PV-barrier mechanism and the strong
KAM stability barrier mechanism act in tandem to strengthen the
barriers near eastward jets; and 2) we have performed a linear
theory Rossby wave analysis of PV-staircase flows that reveals
that Rossby wave critical layers are precluded at the eastward
jets \citep[see also][]{Dunkerton-Scott-07}, which suggest that
the eastward jet barriers may be more robust.

\section{Observational evidence from planetary atmospheres}

In the previous section it was demonstrated that transport
barriers may exist in a PV-conserving flow at locations that do
not coincide with PV-barriers. In this section we discuss
observational evidence that suggests the existence of transport
barriers in the absence of PV-barriers. We consider two
examples: 1) Jupiter's weather layer; and 2) the Earth's
stratosphere. In both cases, conclusions drawn should be
regarded as tentative inasmuch as we do not treat either system
in enough depth to make a definitive statement. In spite of our
incomplete treatment of these topics, we feel that it is
important to point out that, consistent with the theoretical
and numerical results presented in the previous section, there
is observational evidence in planetary atmospheres that
suggests the existence of transport barriers in the absence of
PV-barriers that can be explained by the strong KAM stability
mechanism. Both the examples considered are, in our view,
sufficiently important that the connection between the
observations discussed and the strong KAM stability barrier
mechanism is worthy of a much more thorough investigation.

It is natural to focus on flows consisting of a sequence of
alternating eastward and westward zonal jets because many of
the arguments used in the previous section are then applicable.
In particular, for this class of flows, meridional transport
barriers of the strong KAM stability type are predicted to
occur at the latitudes where $u_0'(y) = 0$. Furthermore, in
alternating zonal jet flows one may anticipate that eastward
and westward jets are associated with large and small
background PV-gradients, respectively. Note, however, that for
the purpose of identifying transport barriers in the absence of
PV-barriers it is not necessary that the background PV
distribution in the flows considered be that of an idealized
PV-staircase. This point is discussed in more detail below. The
following simple scaling argument shows why an alternating
zonal jet flow is well-developed in Jupiter's weather layer but
is only marginally identifiable in the Earth's stratosphere.
These systems are then discussed in turn.

Recall that in a PV-staircase flow the separation between
adjacent eastward jets is $2b = 2\sqrt{3U_{\max}/\beta}$. We
shall assume that this estimate approximately holds for general
midlatitude multiple zonal jet mean flow patterns. Let $r$
denote planetary radius and $\Omega$ planetary rotation rate.
Then $\beta \sim \Omega/r$ and the number of eastward (or
westward) jets one expects to observe in each hemisphere at
midlatitudes (whose extent in latitude is taken here to be half
the equator to pole distance) is approximately $n_\jet \sim
\frac{\pi}{4}r/(2\sqrt{3U_{\max}r/\Omega}) \sim
\frac{1}{4}\sqrt{\Omega/(U_{\max}r)}$. In Jupiter's weather
layer ($U_{\max} \sim 50$ m s$^{-1}$, $r \sim 7 \times 10^4$
km, $\Omega \sim 2\pi/10$ h) $n_\jet \sim 5$, in good agreement
with Fig. \ref{BeltZone}, as described below. In the Earth's
stratosphere ($U_{\max} \sim 50$ m s$^{-1}$, $r \sim 6.4 \times
10^3$ km, $\Omega \sim 2\pi/24$ h) $n_\jet \sim 1$. Thus,
conditions for the formation of a multiple zonal jet mean flow
pattern are only marginally satisfied in the Earth's
stratosphere. In the Earth's troposphere $U_{\max}$ is smaller,
suggesting more favorable conditions, but mountain ranges and
thermal exchange processes between the atmosphere and
irregularly shaped oceans and continents constitute significant
hindrances to the formation of zonal flows. Conditions are
favorable in the Earth's oceans ($U_{\max} \sim 0.5$ m
s$^{-1}$, corresponding to $n_\jet \sim 8$) but there the
presence of lateral boundaries dictates that zonal jets be
embedded in recirculation gyres
\citep{Maximenko-etal-05,Richards-etal-06,Maximenko-etal-08}.

\subsection{Jupiter's weather layer}

\begin{figure}[tb]
\figbox*{}{}{
\includegraphics[width=\textwidth,clip=]{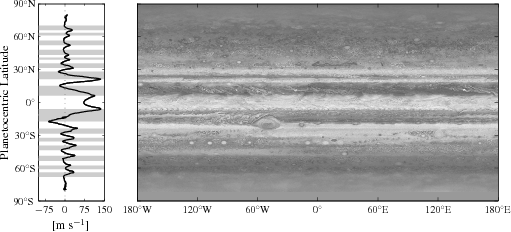}}
\caption{(left panel) Zonally-averaged zonal wind speed as a
function of latitude on Jupiter at the cloud top level as
inferred from images taken in December 2000 by the
\emph{Cassini} spacecraft \citep{Porco-etal-03}. Gray and white
zonal bands indicate belts and zones, respectively. (right
panel) Near instantaneous visible band image of Jupiter
constructed from images taken by the \emph{Cassini} spacecraft
in December 2000. Note that many features of Jupiter's weather
layer that are not seen in this image may be revealed by
radiative transfer analyses outside of the visible band of the
electromagnetic spectrum. NASA image PIA07782 (PIA images are
available at NASA's Planetary Photojournal website
http://photojournal.jpl.nasa.gov).} \label{BeltZone}
\end{figure}

The most striking feature of Jupiter's weather layer
circulation \citep{Porco-etal-03,Vasavada-Showman-05} is that
it is organized in a sequence of alternating eastward and
westward zonal jets whose meridional excursion is very small.
Figure \ref{BeltZone} shows zonally-averaged zonal wind speed
on Jupiter at the cloud top level as a function of latitude.
Regions with dark and light shading in this figure are referred
to as belts and zones, respectively. Belts and zones correspond
to regions in which the background motion is cyclonic $(u'_0(y)
< 0)$ and anticyclonic $(u'_0(y) > 0)$, respectively. The
boundaries between adjacent belts and zones coincide with the
cores of the zonal jets. At these boundaries $u_0'(y) = 0$.
Belts and zones have different radiative transfer properties
(analyses are not limited to the visible band of the
electromagnetic spectrum) which is attributed to differences in
chemical composition
\citep{Carlson-etal-94,Banfield-etal-98,Simon-etal-01,
Irwin-etal-01,Irwin-etal-05}. Assuming that the weather layer
flow is approximately two dimensional and that chemical species
are long-lived, one may infer from the observation that
adjacent belts and zones have different chemical constituents
that there is very little fluid exchange between adjacent belts
and zones, implying that both eastward and westward zonal jets
act as robust meridional transport barriers.

\begin{figure}[tb]
\figbox*{}{}{
\includegraphics[width=\textwidth,clip=]{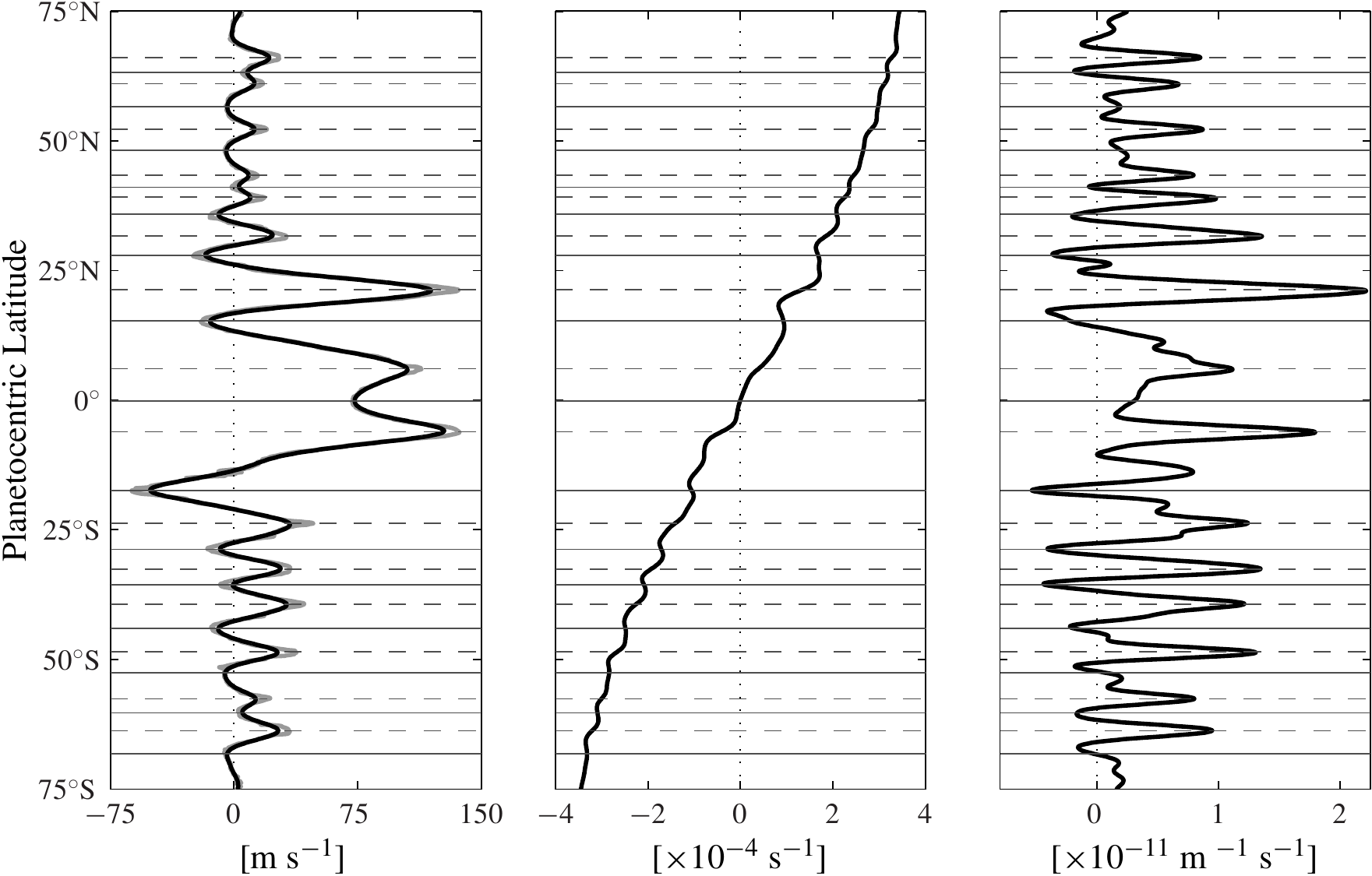}}
\caption{Meridional distribution of zonally-averaged zonal wind
$u_0(y)$ (left panel), zonally-averaged potential vorticity
$q_0(y)$ (center panel), and zonally-averaged potential
vorticity gradient $q_0'(y)$ (right panel) in Jupiter's weather
layer based on \emph{Cassini} data. Both unsmoothed and
smoothed wind profiles are plotted; the smoothed wind profile
was used to compute $q_0(y)$ and $q_0'(y)$. Solid and dashed
horizontal lines are drawn at the latitudes of the cores of
westward and eastward jets, respectively. Note that large
positive values of $q_0'(y)$ coincide with the cores of
eastward jets. A spherical planet (rather than a $\beta$-plane,
which is used elsewhere in this paper) was assumed to compute
the $q_0(y)$ and $q_0'(y)$ structures shown.} \label{uqq_y}
\end{figure}

Figure \ref{uqq_y} shows $u_0(y)$, $q_0(y)$, and $q_0'(y)$ in
Jupiter's weather layer. The same data are shown (but are
plotted differently) in \citet{Read-etal-06a}. The question of
whether a PV-staircase is a useful approximate model of
Jupiter's weather layer has been considered by many authors
\citep{Read-etal-06a,Marcus-Lee-98,Peltier-Stunhe-02,Dowling-95a}.
For our purposes the answer to this question is not critical.
Our focus is on identifying transport barriers that cannot be
explained by the PV-barrier mechanism. Figure \ref{uqq_y} shows
that while all eastward jets are associated with large
meridional PV-gradients, most of the westward jets are
associated with small meridional PV-gradients. (Here and above
we are using the term westward jet somewhat loosely to include
minima of $u_0(y)$, even when $u_0 > 0$ at the minimum.) In
other words, most of the transport barriers near the cores of
the westward jets cannot be explained by the PV-barrier
mechanism. But all of the belt-zone boundaries---the apparent
meridional transport barriers---coincide with latitudes at
which $u_0'(y) = 0$, consistent with the strong KAM stability
barrier mechanism. Thus all of the apparent transport barriers
can be explained by the strong KAM stability barrier mechanism.
Note, however, that there is no apparent barrier on the
equator, where $u_0'(y) = 0$. This is probably due to a
combination of anomalous equatorial dynamics
\citep{Heimpel-etal-05} and the manner by which chemical
constituents are pumped into the near-equatorial weather layer.

Some caveats relating to our interpretation of observations
from Jupiter should be emphasized, however. First, we have
assumed that the weather layer flow is nearly two dimensional
and horizontally nondivergent, being only weakly forced by
convection. Although these assumptions are generally accepted
\citep[see, for example,][]{Vasavada-Showman-05}, it should be
noted that our explanation of the apparent transport barrier
rests on their validity. A second assumption that we have made
is that chemical species in Jupiter's weather layer are
long-lived. Another possible explanation of the apparent
transport barriers between adjacent belts and zones is that
chemical species are short-lived, being continuously pumped
into the weather layer by convective overturning
\citep{Ingersoll-etal-00,Showman-dePater-05}. We cannot rule
out this possibility. Our argument shows, however, that the
apparent lack of fluid exchange between adjacent belts and
zones \emph{can} be explained using dynamical arguments.

\subsection{The Earth's stratosphere}

\begin{figure}[tb]
\figbox*{}{}{
\includegraphics[width=.8\textwidth,clip=]{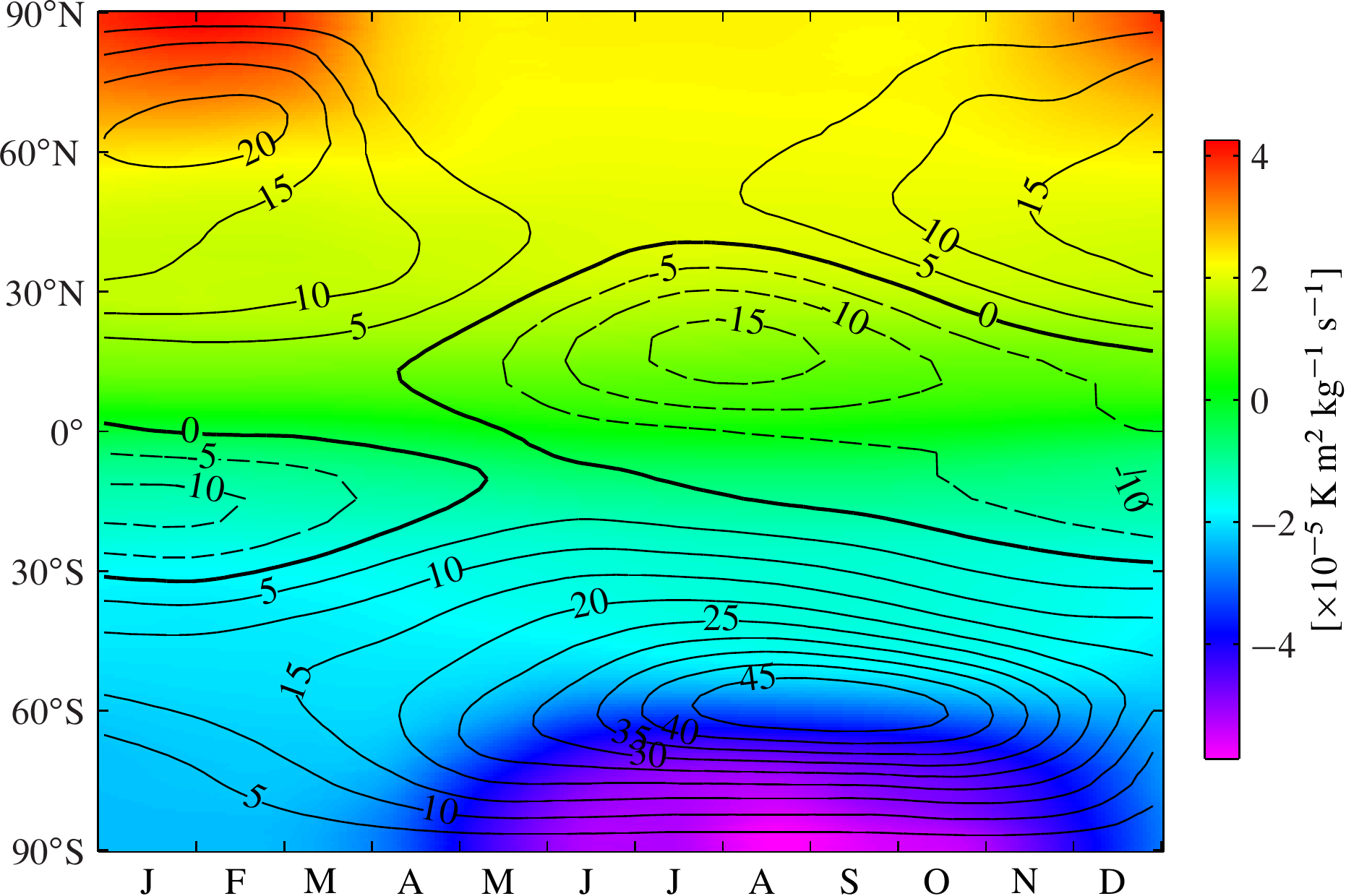}}
\caption{Seven-year (1992--1998) monthly average of
zonally-averaged zonal wind and potential vorticity on the 460
K isentropic surface based on the ERA-40 reanalysis product.
Contours show zonal winds in meters per second. Solid and
dashed contours represent eastward and westward flow,
respectively. Colors indicate Ertel's potential vorticity.}
\label{era40}
\end{figure}

The simple scaling argument given above predicts that
conditions for the formation of a stable alternating multiple
jet zonal mean flow pattern are only marginally satisfied in
the Earth's stratosphere. In qualitative agreement with this
prediction, in each hemisphere there is one readily
identifiable eastward zonal jet and one westward jet, and the
appearance of these jets is seasonal \citep[see, for
example,][]{Andrews-etal-87}. The stronger jets are the high
latitude eastward polar night jets which appear in the winter
hemisphere. (The austral polar night jet is particularly
strong, persisting throughout most of the stratosphere during
the austral fall, winter and spring.) The westward jets are
present in the subtropics throughout most of the stratosphere
during the summer months in each hemisphere. The eastward polar
night jets, especially in the southern hemisphere, act as
transport barriers and are associated with strong PV-gradients.
Because the focus in the present study is on identifying
transport barriers in the absence of PV-barriers, these jets
are not of interest here. In contrast, the westward subtropical
jets in the summer hemisphere are very much of interest because
these are not associated with a strong PV-gradient. The
properties just described are illustrated in Fig. \ref{era40}.
That figure shows a 7-year (1992--1998) monthly average of
zonally-averaged potential vorticity (in color) and zonal wind
(as contours) on the 460 K isentropic surface (which lies in
the lower stratosphere) based on the European Center for
medium-Range Weather Forecasts (ECMWF) 40-year Re-Analysis
(ERA-40) product. Both the eastward polar night jets and the
westward subtropical jets are readily identifiable. As we have
noted, the eastward winter hemisphere polar night jets are
associated with strong meridional PV-gradients while the
westward subtropical jets are associated with nearly homogenous
PV-distributions. The westward subtropical jets have the
dynamical properties that we seek---zonal jets in the absence
of strong PV-gradients. We note also that, consistent with
arguments made originally by \citet{Charney-Drazin-61}, Rossby
wave perturbations to the background in these regions are weak.
From the standpoint of applicability of the strong KAM
stability argument, weak perturbations are advantageous.

We turn our attention now to the question of whether the cores
of these jets serve as robust meridional transport barriers.
Studies of stratospheric transport based on effective
diffusivity have been carried out by \citet{Allen-Nakamura-01}
and \citet{Haynes-Shuckburgh-00}. The effective diffusivity is
large (small) in regions where fluid is well (poorly) mixed.
Fluid in the vicinity of a transport barrier is poorly mixed;
these regions are thus characterized by a small effective
diffusivity. Both stratospheric effective diffusivity analyses
reveal that the westward subtropical jet in the summer
hemisphere coincides with a region of anomalously low effective
diffusivity; see plates 1 and 4 in \citet{Allen-Nakamura-01}
and 1 through 4 in \citet{Haynes-Shuckburgh-00}. This suggests
that these westward jets act as meridional transport barriers.
Previous work by \citet{Waugh-96} and \citet{Chen-etal-94} had
focused on this ``subtropical barrier.'' Indeed, this barrier
comprises a critical element of the ``tropical pipe'' model
\citep{Plumb-96} of stratospheric transport. Observational
evidence that suggests the presence of subtropical transport
barriers is presented in \citet{Grant-etal-96},
\citet{Mote-etal-98}, \citet{Minschwaner-etal-96},
\citet{Trepte-Hitchman-92} and \citet{Trepte-etal-93}.
\citet{Shepherd-07} provides a recent review of stratospheric
transport, including a discussion of subtropical transport
barriers.

The evidence that we have pointed out strongly suggests that:
1) the stratospheric subtropical barrier is a robust meridional
transport barrier that coincides with the core of a westward
jet; 2) the associated meridional PV-gradient is very small so
this barrier cannot be explained by the PV-barrier mechanism;
and 3) because $u_0'(y) = 0$ at this barrier, the barrier is
predicted by the strong KAM stability barrier mechanism. To
test these tentative conclusions more rigorously, a study based
on realistic synoptic winds that tracks both potential
vorticity and tracer distributions should be conducted.

\section{Summary and discussion}

In the first part of this paper we presented numerical
simulations of passive tracer transport in a perturbed
PV-staircase flow and showed that both eastward and westward
jets in this flow act as meridional transport barriers. The
surprise is that westward jets, where the background PV
gradient vanishes, act as transport barriers. This behavior was
explained as being a consequence of the strong KAM stability
barrier mechanism.

We then briefly discussed the applicability of the strong KAM
stability mechanism to explaining observations of Jupiter's
weather layer and the Earth's subtropical stratosphere. In both
of these systems westward jets are present that appear to act
as robust meridional transport barriers in the absence of a
background meridional PV-barrier. These barriers are predicted
by the strong KAM stability mechanism. In both cases the
evidence presented should be regarded as suggestive. More
thorough investigations of both problems is recommended.

The principal weakness of our explanation of the apparent lack
of fluid exchange between adjacent belts and zones in Jupiter's
midlatitude weather layer is that we cannot exclude the
possibility that the observed chemical composition differences
in adjacent belts and zones are caused by a combination of
strong convective overturning and short-lived chemical species.
In spite of this caveat, it is important to emphasize that we
have shown that maintenance of the apparent chemical
composition differences between adjacent belts and zones
\emph{can} be explained using a dynamical argument (as opposed
to a chemistry-based argument) in which the weather layer flow
is only weakly convectively forced.

The principal weaknesses in our discussion of the Earth's
subtropical stratospheric transport barrier were that all of
the properties noted were based on averaged winds rather than
synoptic winds and that tracer transport and potential
vorticity distributions were not estimated in a way that was
guaranteed to be self consistent. It should not be difficult to
overcome these shortcomings using model-based synoptic winds.

\acknowledgements We thank T. Dowling and R. Morales-Juberias
for providing the data used to construct Figs. \ref{BeltZone}
and \ref{uqq_y}. Comments on the manuscript by anonymous
reviewers, T. Shepherd, T. {\"O}zg{\"o}kmen, and J. Willemsen
are sincerely appreciated. The ECMWF ERA-40 data used in this
study were obtained from the ECMWF data server. Support for
this work was provided by the U.S. National Science Foundation
under grants CMG0417425 and OCE0648284.

\bibliographystyle{ametsoc}

\end{article}

\end{document}